\documentclass[referee,useAMS,usenatbib,usegraphicx]{mn2e}
\title[Sausage oscillations in cool postflare loop]{Observation of multiple sausage oscillations in cool postflare loop}
\author[Srivastava et al.]{A.K.~Srivastava$^{1}$\thanks{E-mail:
aks@aries.ernet.in}, T.V.~Zaqarashvili$^{2}$, W.~Uddin$^{1}$, B.N.~Dwivedi$^{3}$, Pankaj Kumar $^{1}$\\\footnotemark[1]\thanks{Send off-print
request to Dr. A.K. Srivastava, ARIES, Manora Peak, Nainital-263 129, India.}\\
  $^1$Aryabhatta Research Institute of Observational Sciences (ARIES), Nainital-263129, India\\
  $^2$Abastumani Astrophysical Observatory at I. Chavchavadze State University, Al Kazbegi ave. 2a, 0160 Tbilisi,Georgia\\
  $^3$Department of Applied Physics, Institute of Technology, Banaras Hindu University, Varanasi-221005, India}

\begin{document}

\date{Accepted . Received }

\pagerange{\pageref{firstpage}--\pageref{lastpage}} \pubyear{2008}

\maketitle

\label{firstpage}

\begin{abstract}
Using simultaneous high spatial (1.3 arc sec) and temporal (5 and 10 s)
resolution H-alpha observations from the 15 cm Solar Tower Telescope
at ARIES, we study the oscillations in the relative intensity to
explore the possibility of sausage oscillations in the chromospheric
cool postflare loop. We use standard wavelet tool, and find the
oscillation period of $\approx$ 587 s near the loop apex, and
$\approx$ 349 s  near the footpoint. We suggest that the
oscillations represent the fundamental and the first harmonics of
fast sausage waves in the cool postflare loop. Based on the period ratio
$P_1/P_2 \sim$ 1.68, we estimate the density scale height in the
loop as $\sim$ 17 Mm. This value is much higher than the
equilibrium scale height corresponding to H-alpha temperature, which
probably indicates that the cool postflare loop is not in
hydrostatic equilibrium. Seismologically estimated Alfv\'en speed
outside the loop is $\sim$ 300-330 km/s. The observation of multiple
oscillations may play a crucial role in understanding the dynamics
of lower solar atmosphere, complementing such oscillations already
reported in the upper solar atmosphere (e.g., hot flaring
loops).
\end{abstract}


\section{Introduction}
The coupling of complex magnetic field and plasma generates variety
of magnetohydrodynamic (MHD) waves and oscillations in various solar
structures. These MHD waves and oscillations are one of the
important candidates for coronal heating and solar wind
acceleration. The idea of exploiting observed oscillations as a
diagnostic tool for determining the physical conditions of the
coronal plasma was first suggested by Roberts et al. (1984). Until
recently, the application of this idea has been marked by the lack
of high-quality observations of coronal oscillations. However, this
situation has changed dramatically, especially due to space-based
observations by the Solar and Heliospheric Observatory (SOHO), the
Transition Region and Coronal Explorer (TRACE), and most recently
with the high resolution spectra from the Hinode spacecraft. The
fast kink wave is most frequently observed mode as it can be
directly detected by periodic spatial displacement of coronal loop
axis (Nakariakov et al. 1999; Aschwanden et al. 1999; Wang \&
Solanki 2004). Using temporal series image data from CDS/SOHO,
recently O'Shea et al. (2007) have found the first evidence of
fast-kink standing oscillations in the cool transition region loops
as well. On the other hand, the fast MHD sausage wave causes the
variation of pressure and magnetic field in a coronal loop and
therefore can be observed as intensity oscillations or periodic
modulation of coronal radio emission (Nakariakov et al. 2003).
Recently, Srivastava  et al. (2008) have reported the signature of
the leakage of chromospheric magnetoacoustic oscillations into the
corona near the south pole. These observations provide a basis for
the estimation of coronal plasma properties (Nakariakov \& Ofman
2001).

Recently, Verwichte et al. (2004) have detected interesting phenomenon of
simultaneous existence of fundamental and first harmonics of fast
kink oscillations (see also De Moortel \& Brady 2007; Van
Doorsselaere et al. 2007). However, the ratio between the periods of
fundamental and first harmonics $P_1/P_2$ was significantly shifted
from 2, which later was explained as a result of longitudinal
density stratification in the loop (Andries et al. 2005; McEwan et
al. 2006). The rate of the shift allows to estimate the density
scale height in coronal loops, which can be few times larger
compared to its hydrostatic value (Aschwanden et al. 2000).

In this paper, we report the first evidence of fundamental and first
harmonics of fast sausage oscillations in cool postflare loop. We
use high spatial (1.3 arc sec) and temporal (5 and 10 s) resolution
H-alpha observations from the 15 cm Solar Tower Telescope at ARIES,
Nainital, India, to study the oscillations in the relative
intensity. Using the standard wavelet tool, we find multiple
oscillation periods at different parts of the loop, which are
interpreted as the fundamental and first harmonics of fast sausage
mode. In section 2, we describe the observations and data reduction.
We describe the wavelet analysis in section 3. In section 4, we
present our theoretical model and its link to coronal seismology. We
present results and discussion in the last section.

\section{Observations and data reduction \label{sec:obs}}
The observations of post flare loops have been carried out
with 15 cm, f/15 Coud\'e Solar Tower Telescope at ARIES, Nainital, India, equipped with 
Bernhard Hale H-alpha filter ($\lambda$ = 6563 \AA\  and P.B. 0.5/0.7 \AA\ ),
and PXL Photometric CCD camera. The filter is a birefringent Lyot type filter, tunable
$\pm$ 1 \AA\ from central wavelength (6563 \AA\ ) with the step
of 0.1 \AA\ . During our observations, we set it at central wavelength
6563 \AA\ with passband 0.5 \AA\ . The Image size has been
enlarged by a factor of two using a Barlow lens. The 512 $\times$
512 pixels, 12-bit frame transfer CCD camera has a square pixel of
15 {\it {$\mu$}m$^{2}$} corresponding to a 0.65 arc sec pixel size.
The spatial resolution of our observations is 1.30 arc sec.
The read out noise for the system is 31 e$^{-}$ with a gain of 44.1 e$^{-}$/ADU. Dark
current of the camera is 56 e$^{-} s^{-1}$. The dark current
is integrated over the 512 x 512 pixels. The camera
controller of the system has a variable read-out rate from 0.5 to 2
MHz.The variable read out rate of our CCD provides us the
facility of fast imaging of the flares at different rates. In these
observations, we have used a constant read out rate of 2 MHz. The
The CCD chip (EEV 37) is cooled up to -25$^{o}$ C using liquid
cooling system.
%%%%%%%%%%%%%%%%%%%%%%%%%%%%%%%%FIG 1%%%%%%%%%%%%%%%%%%%%%%%%%%%%%%%%%%%%%%%%%%%%%%%%%%
%\begin{verbatim}
\begin{figure}
\centering
%\vspace*{-0.3cm}
%\vspace*{174pt}
\includegraphics[width=60mm,angle=0]{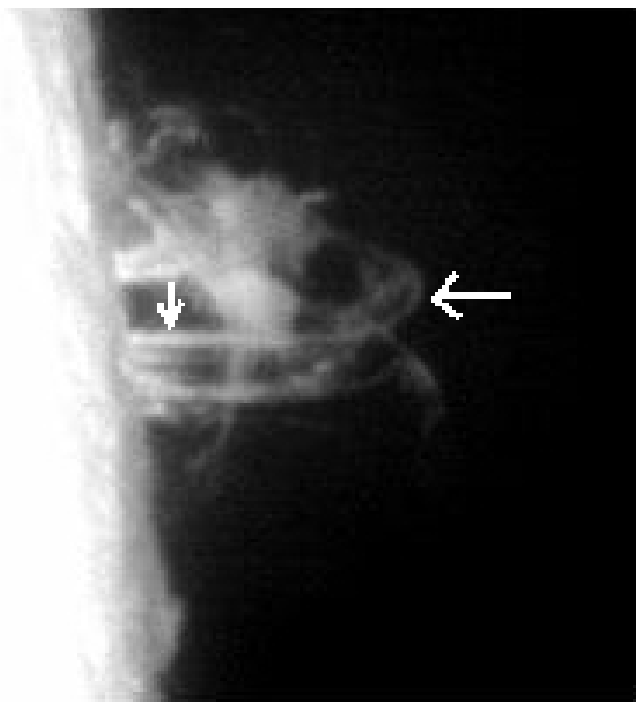} \includegraphics[width=59mm,angle=0]{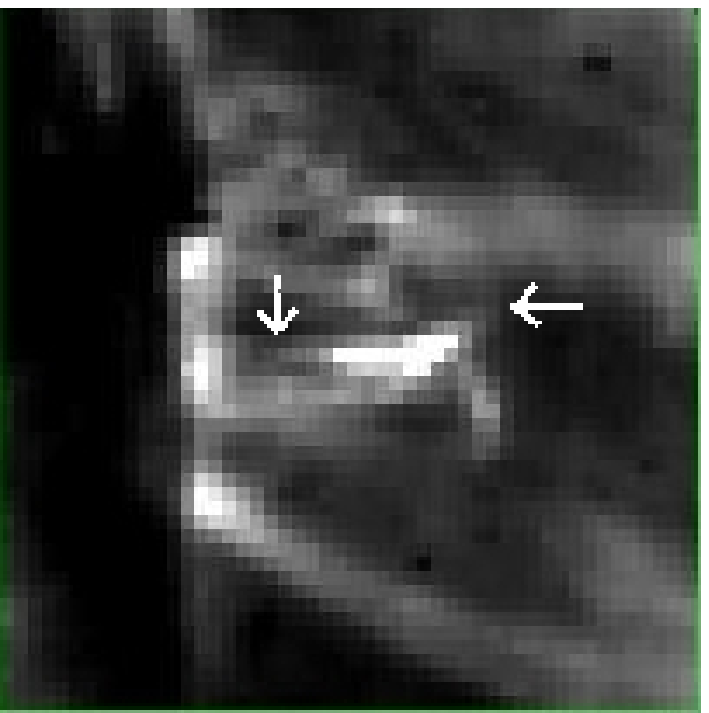}
%\includegraphics{loopc.ps}
%\vspace*{-0.5cm}
\caption{The H-alpha post flare loops as observed by 15 cm Solar
Tower Telescope at ARIES, Nainital, India. The FOV is 200 pixel
$\times$ 200 pixel (or 130 arc sec $\times$ 130 arc sec) (left panel).The aligned
SOHO/EIT Fe IX/X 171 \AA\ image of the same loop system is presented in right panel.}
\label{fig1}
\end{figure}
%\end{verbatim}
%%%%%%%%%%%%%%%%%%%%%%%%%%%%%%%%%%%%%%%%%%%%%%%%%%%%%%%%%%%%%%%%%%%%%%%%%%%%%%%

We analyze the postflare loop system between 01:00:52 U.T.
and 01:58:28 UT on May 2, 2001, at a cadence of 5 s and 10 s with
an exposure time of 30 ms. The post flare loop system was associated with
M1.8 class limb flare which occured in NOAA active region AR 9433 (N15, W88).
Since, the temporal evolution/changes were very fast during the
onset of post flare loop system, we took fast sequence of images with 5 s cadence.
As time progresses in the gradual/decay phase, the temporal changes
become slow compared to initial phase of post flare loop system. So,
we took the sequence of images with 10 s cadence. Hence, there
are two cadences, initially 5 s and  later on 10 s. We have
started our fast imaging with 5 s cadence on 00:52:11 UT. In the
beginning of our observations, the flare was in decay phase and was
enveloped by complex post flare loop system. Hence, we were unable
to select the distinct loop initially. The post flare loop system
became more relaxed and distinctly visible around
01:00:52 UT, then we have selected images of a
clearly visible loop for our study after 01:00:52 UT. In our analysis, first 92 frames were with 5 s
cadence, while rest of the frames were with 10 s cadence. The total number
of the frames used in this study is 390. We find that the post flare loop
system shows the similar structure/evolution in SOHO/EIT (171 \AA\ , 195 \AA\ , 284 \AA\
 , 304 \AA\ ) and H-alpha observations.The post flare loop system which was visible at higher temperature
(6.0$\times$10$^{4}$ K--3.0$\times$10$^{6}$ K) with EIT, has also seen in H-alpha
at the lower temperature ($\approx$ 10$^{4}$ K) after its cooling. In Fig 1 (top panel), the image of
the post flare loop system in high resolution H-alpha has been shown,
while the same loop system observed with SOHO/EIT Fe IX/X 171 \AA\ has been
shown in the bottom panel.Both images show the clear shape of the
dynamic post flare loop that we have selected. The loop top and foot
point of this prominent loop marked by arrows in these two images. Since the spatial
resolution of our H-alpha observations is  high in comparison
to the SOHO/EIT, hence we can not see the features in EIT as sharp
as in H-alpha image. 

	We have estimated the length of the selected 
loop in H-alpha and SOHO/EIT Fe IX/X 171 \AA\ images. We find the lengths
$\sim$ 97 Mm and $\sim$ 100 Mm in the H-alpha and EUV images respectively.
The estimated loop lengths at two wavelengths are almost equal which clearly
shows the presence of the same loop at different temperatures.
The lower atmosphere may not necessarily be an
optically thin atmosphere, and many bright structures cross to
each other in the field of view. However, the chromospheric post
flare loop system is very complex and dynamic in itself, instead of
simpler and static. The features are temporally changing very fast
in the observations. So, we carefully selected the image frames of
the clearly visible loop in order to study the oscillations. The sky
condition was very good during the observations. Several dark and flat field
images have also been taken to calibrate the H-alpha images. Using IRAF and SSWIDL softwares,
the image processing has been performed. The details of the instrument can be obtained
from Ali et al. (2007) and  Joshi et al. (2003).

\section{Wavelet Analysis}

We use the Morlet wavelet tool to produce the power spectrum of
oscillations in H-alpha. The Morlet wavelet suffers from the edge
effect of time series data. This effect is significant in regions
defined as cone of influence (COI). The details of wavelet
procedure, its noise filtering, COI effects etc. are given in
Torrence \& Compo (1998). Randomisation technique evaluates the peak
power in the global wavelet spectrum, which is just the average peak
power over time and similar to a smoothed Fourier power spectrum.
This technique compares it to the peak powers evaluated from the
$n!$ equally likely permutations of the time series data, assuming
that n values of measured intensities are independent of n measured
times if there is no periodic signal. The proportion of permutations
which gives the value greater or equal to the original peak power of
the time series, will provide the probability of no periodic
component (p). The percentage probability
%%%%%%%%%%%%%%%%%%%%%%%%%%%%%%%%FIG 2%%%%%%%%%%%%%%%%%%%%%%%%%%%%%%%%%%%%%%%%%%%%%%%%%%
%\begin{verbatim}
\begin{figure}
%\vspace*{-0.3cm}
%\vspace*{174pt}
\includegraphics[width=70mm,angle=90]{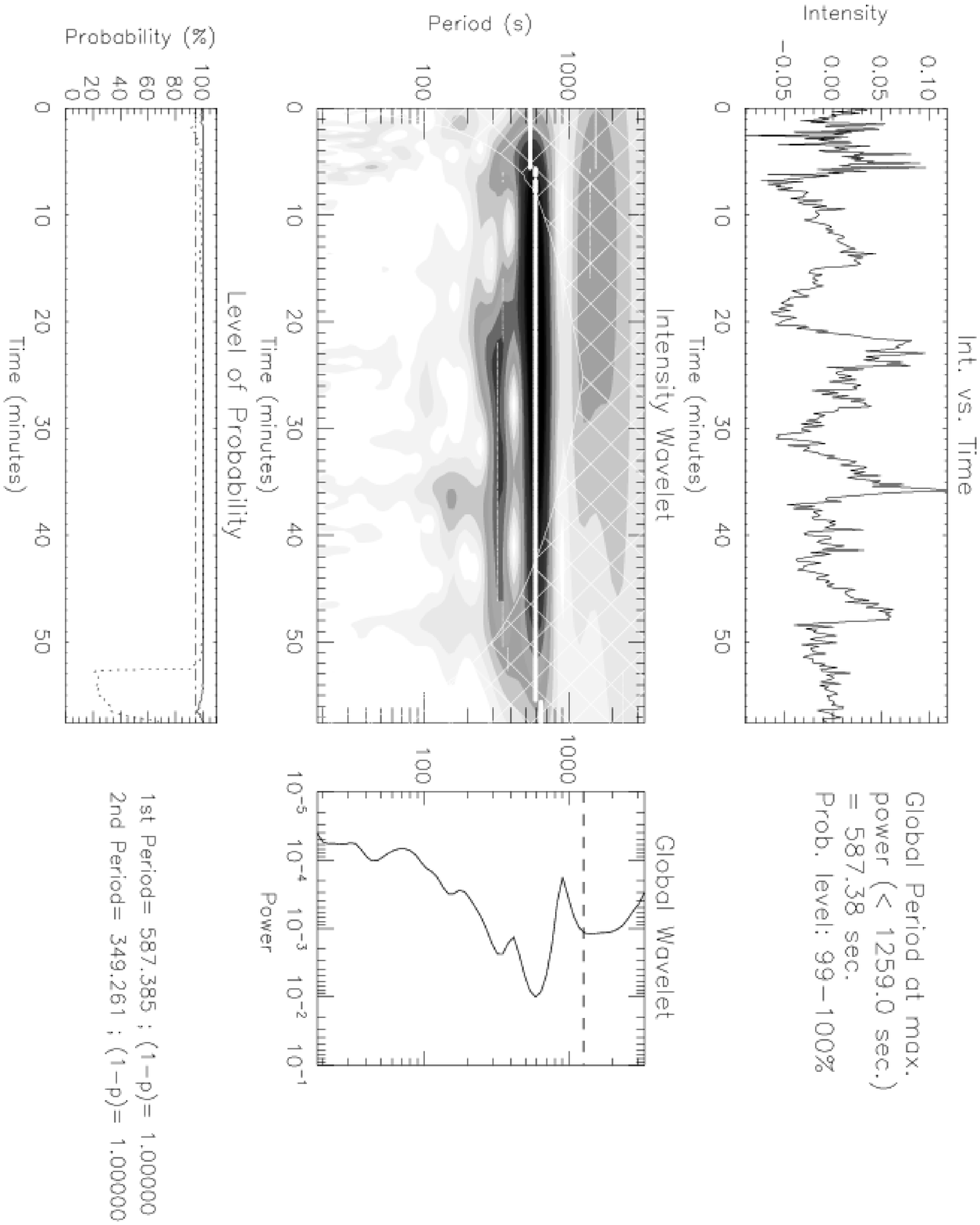}
%\includegraphics{new_wlt_apex.eps}
%\vspace*{-0.5cm}
\caption{The wavelet result for H-alpha 6563 \AA\ line near the loop
apex. The top panel shows the variation of intensity, the wavelet
power spectrum is given in the middle panel, and the probability in
the bottom panel. The light curve is from the loop apex {\it (X,Y)}
= (133th pixel, 117th pixel), and smoothed by window width 60.}
\label{fig2}
\end{figure}
%\end{verbatim}
%%%%%%%%%%%%%%%%%%%%%%%%%%%%%%%%%%%%%%%%%%%%%%%%%%%%%%%%%%%%%%%%%%%%%%%%%%%%%%%
of periodic components presented in the data will be $(1-p)\times
100$, and 95 \% is the lowest acceptable probability for real
oscillations. We set 200 permutations for the reliable estimation of
p, and hence the probability of real oscillations. The details of
randomisation technique to obtain the statistically significant real
oscillation periods are given by Nemec \& Nemec (1985) and O'Shea et
al. (2001). We do not remove any upper/longer period intervals and
associated powers of our time series data during wavelet analysis.
We choose the 'running average' option of O'Shea's wavelet tool, and
smooth the original signal by window width 60. Average smoothing
process is used to reduce the noise in the original signal in order
to get a real periodicity, and it is based on the low pass filtering
methods. The 'Running Average' process smooth the original
signal by the defined scalar width. This process use the 'SMOOTH'
subroutine available with IDL tool kit. The SMOOTH function returns
a copy of array smoothed with a boxcar average of the specified
width. The result has the same type and dimensions as array. The
fitted signal is then subtracted from the original signal, and gives
the resultant signal for the wavelet analysis. The maximum allowed
period from COI, where edge effect is more effective, is 1259 s.
Hence, the power reduces substantially beyond this threshold. In our
wavelet analysis, we only consider the power peaks and corresponding
real periods  below this threshold. We performed the wavelet
analysis at different parts (near the apex and foot point) of the
selected loop. We have chosen a location of {\it (X,Y)} = (133th
pixel, 117th pixel) near the apex, where the dominant periodicity is
$\approx$ 587 s with the probability of $\approx$ 99 \%--100 \%
(Fig. 2). We have again chosen a location of {\it (X,Y)} = (58th
pixel, 104th pixel) near the foot point of the same loop, where the
dominant oscillation period is $\approx$ 349 s with the probability
of $\approx$ 99 \%--100 \% (Fig. 3). It should be noted that we have
chosen a box of width '4' at these locations to extract the light
curves with good signal-to-noise (S/N) ratios. Boxes with
the width '4' have been chosen near the loop-top and foot point with
respect to their reference coordinates as mentioned in the figure
captions of the wavelet diagrams. We extract maximum intensity
from the chosen box.  Our box of width '4' fits the loop dimension
best. However, we have checked the results at other widths, e.g.,
'2', '3' , and '6'. The boxes with widths '2' and '3' still remain
inside the loop dimension. However, the box with width '6' crosses
the loop, which may not be appropriate especially near the limb
where the background emissions and structures may be more effective
compared to far off the limb.

From the boxes of width '2' and '3', we get the same global
periodicities at maximum power as we have obtained in the case of
box width '4'. However, the power distribution in the intensity
wavelet is slightly different for each case. This
is obvious because we extract lower intensity with smaller box and
higher intensity with larger box. We get a periodicity of $\sim$ 493
s and $\sim$ 587 s respectively near the loop
foot point and loop apex using the box of width '6'. The
periodicity at apex is similar to the previous findings because the
limb effects are not much far off the limb. The difference in the
periodicity near loop foot point shows the out side effects near the
limb, which came into action due to the larger box size compared to
the loop dimension. However, the similar periodicities ($\sim$ 349 near the foot point and
$\sim$ 587 near the apex) with well
fitted or smaller boxes (of widths '2', '3', and '4'), clearly show that we are inside a well
isolated loop structure. However, we should take care of the
dimension of the box during wavelet analysis especially near the
limb.Thus, the selected loop intensities show the oscillations
with significantly different periods near the apex ($\approx$ 587 s)
and footpoint ($\approx$ 349 s) respectively. These observation may
lead an interesting consequences for loop oscillation phenomena.

%%%%%%%%%%%%%%%%%%%%%%%%%%%%%%%%FIG 3%%%%%%%%%%%%%%%%%%%%%%%%%%%%%%%%%%%%%%%%%%%%%%%%%%
%\begin{verbatim}
\begin{figure}
%\vspace*{-0.3cm}
%\vspace*{174pt}
\includegraphics[width=70mm,angle=90]{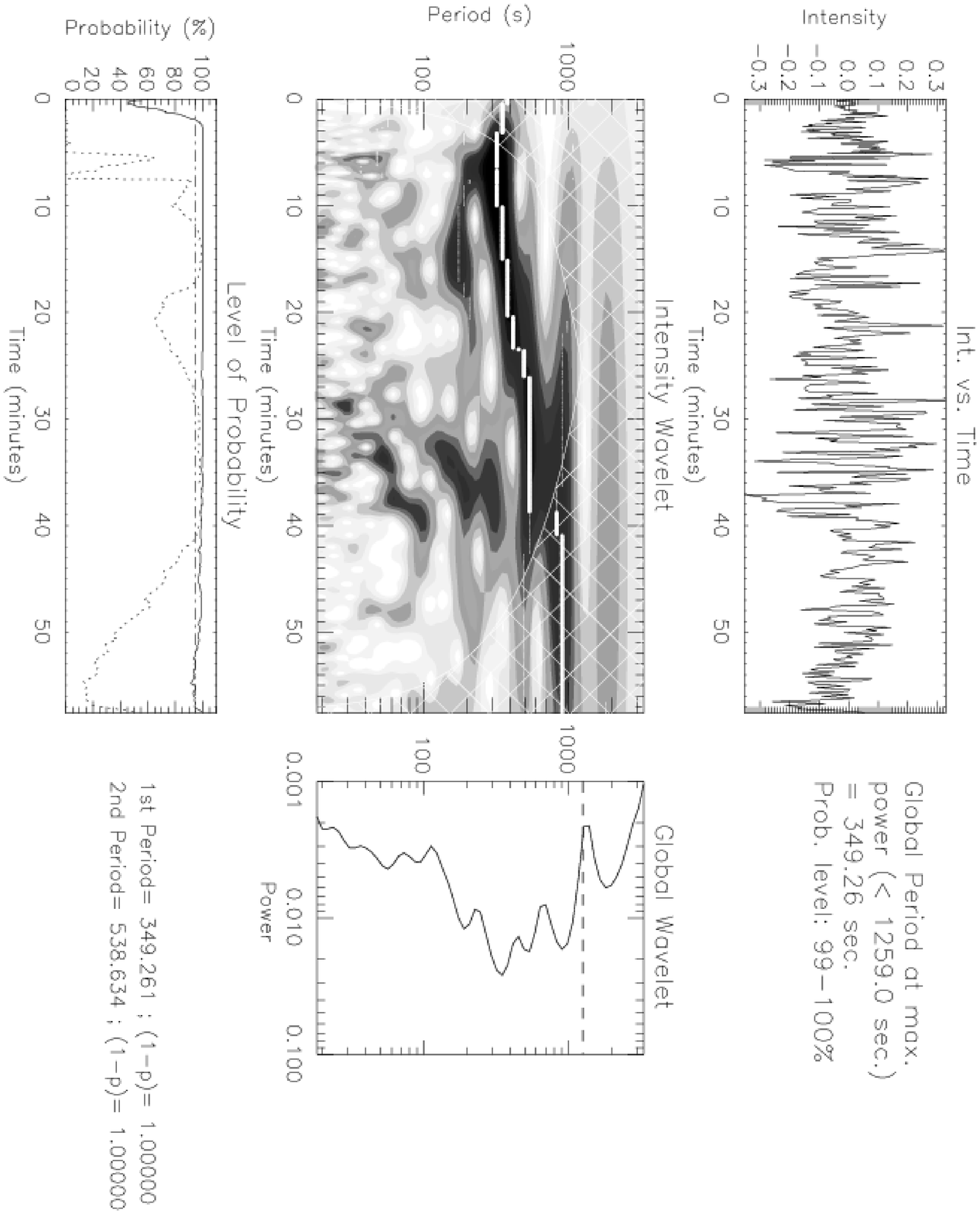}
%\includegraphics{new_wlt_fp.eps}
%\vspace*{-0.5cm}
\caption{The wavelet result for H-alpha 6563 \AA\ line near the loop
foot point. The top panel shows the variation of intensity, the
wavelet power spectrum is given in the middle panel, and the
probability in the bottom panel. The light curve is from loop foot
point {\it (X,Y)} = (58th pixel, 104th pixel), and smoothed by
window width 60.} \label{fig3}
\end{figure}
%\end{verbatim}
%%%%%%%%%%%%%%%%%%%%%%%%%%%%%%%%%%%%%%%%%%%%%%%%%%%%%%%%%%%%%%%%%%%%%%%%%%%%%%%
\section{A Theoretical Interpretation \label{sec:theory}}

The intensity oscillation with the period of $P_1 \approx$ 587 s
likely represents the fundamental harmonic of either fast sausage
(Nakariakov et al. 2003) or slow magnetoacoustic (Nakariakov et al.
2004) waves. The loop length is estimated as $L \approx$ 97 Mm with
corresponding phase speed as $2L/P_1 \sim$ 330 km/s. The estimated
phase speed is much higher than the sound speed corresponding to
cool H-alpha line. Therefore, we suggest that the oscillation is due
to a fast sausage mode.

%{\bf Using cold plasma approximation, the dispersion relation for
%sausage oscillations in straight coronal loop with radius $a$,
%density $\rho_0$ and magnetic field $B_0$ is (Edwin and Roberts,
%1983)
%\begin{equation}
%\rho_0(k^2 v^2_A -\omega^2)m_e{{K_1(m_e a)}\over {K_2(m_e
%a)}}=\rho_e(k^2 v^2_{Ae} -\omega^2)n_0{{J_1(n_0 a)}\over {J_2(n_0
%a)}},
%\end{equation}
%where $k$ is the longitudinal wave number, $\omega$ is the wave
%frequency, $\rho_e$ and $B_e$ are the plasma density and the
%magnetic field outside the loop, $v_A=B_0/\sqrt{4\pi \rho_0}$ and
%$v_{Ae}=B_e/\sqrt{4\pi \rho_e}$ are Alfv\'en speeds inside and
%outside, $m_e=\sqrt {k^2 -\omega^2/v^2_{Ae}}$, $n_0=\sqrt
%{\omega^2/v^2_{A} - k^2}$ and $J_n$ (and $K_n$) are the Bessel
%functions (and modified Bessel functions) of order $n$.}

However, the dispersion relation of fast sausage waves in straight
coronal loop includes both trapped and leaky modes depending on the
loop parameters (Edwin \& Roberts 1983; Nakariakov et al. 2003;
Aschwanden et al. 2004). Due to low sound speed, we can easily use
the cold plasma approximation. Hence, the cut-off wavenumber $k_c$ is
(Edwin and Roberts 1983; Roberts et al. 1984)
\begin{equation}
k=k_c=\left [ {v^2_A \over {v^2_{Ae}-v^2_A}} \right ]^{1/2}{j_0\over
a},
\end{equation}
where $v_A$ and $v_{Ae}$ are Alfv\'en speeds inside and outside the
loop, $a$ is the loop radius and $j_0=2.4$ is the first zero of the
Bessel function $J_0$. The modes with $k>k_c$ are trapped in the
loop, while the modes with $k<k_c$ are leaky. We may assume $B_0=B_e$,
then the cut-off wavenumber is (Aschwanden et
al. 2004)
\begin{equation}
k_c=\left [ {1 \over {{n_0/n_e}-1}} \right ]^{1/2}{j_0\over a},
\end{equation}
where $n_0/n_e$ is the electron density ratio inside and outside the
loop. The width of selected loop can be roughly estimated as $2a
\approx$ 6 Mm. Hence, the trapped global sausage mode can be
realized only if the density ratio is
\begin{equation}
{n_0\over n_e} > \left ({j_0\over \pi}\right )^2 \left ({L\over
a}\right )^2 \approx 600.
\end{equation}
Thus, only very dense loop can support non-leaky global sausage mode
in the estimated length and width (Nakariakov et al. 2003;
Aschwanden et al. 2004). The postflare loops usually have a very
high density contrast of the order of $10^2$--$10^3$. The selected
loop is cool post flare one, therefore the estimated density ratio
falls in the expected range. Thus, the trapped sausage mode may
occur in the loop, however, wave leakage can not be ruled out.

The fundamental harmonic of sausage oscillation has pressure
antinode at the loop apex (upper panel in Fig. 4). Therefore, it
modifies the density leading to the intensity oscillation.
However, it has pressure node at the footpoints, and should not
lead to remarkable intensity oscillation. The observations show
strong oscillations at the loop apex with the period of $\approx$
587 s and almost no indication of the oscillation with this period
near foot point (see Fig. 2). Therefore, the oscillation with
$\approx$ 587 s is probably due to the fundamental mode of sausage
oscillations.

On the other hand, the first harmonic of sausage oscillation has
pressure node at the loop apex and antinodes at the mid parts (lower
panel in Fig. 4). Therefore, the first harmonic should show the
intensity oscillation near the foot point and almost no oscillation at
the loop apex (see Fig. 3). Indeed, the oscillation with the period
of $\approx$ 349 s has high probability near foot point and no
significant probability near the apex. Therefore, we interpret the
oscillation with the period of $\approx$ 349 s as the first harmonic
of fast sausage waves in the coronal loop. In the curved geometry,
this mode is identical to sausage swaying mode (D\'iaz et al. 2006).

It must be mentioned that 349 s oscillation may be a consequence of
nonlinear excitation from the fundamental harmonic, however, this
possibility can be ruled out by two reasons: First, the amplitude of
oscillation is not very strong and leaving a little space for
nonlinear interaction; Secondly, the period of the first harmonic is
not exactly twice of the period of fundamental mode, which was
expected from the resonance condition.

%%%%%%%%%%%%%%%%%%%%%%%%%%%%%%%%FIG 4%%%%%%%%%%%%%%%%%%%%%%%%%%%%%%%%%%%%%%%%%%%%%%%%%%
%\begin{verbatim}
\begin{figure}
\centering
%\vspace*{-0.3cm}
%\vspace*{174pt}
\includegraphics[width=60mm,angle=0]{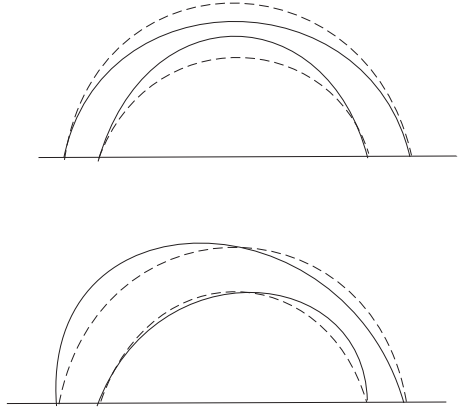}
%\includegraphics{fig6.eps}
%\vspace*{-0.5cm}
\caption{The fundamental (upper panel) and its first (bottom panel)
harmonics of sausage oscillations in coronal loops.} \label{fig4}
\end{figure}
%\end{verbatim}
%%%%%%%%%%%%%%%%%%%%%%%%%%%%%%%%%%%%%%%%%%%%%%%%%%%%%%%%%%%%%%%%%%%%%%%%%%%%%%%

As a consequence of above discussions, we suggest for the first
observational evidence of the fundamental and its first harmonics of
fast sausage oscillations in cool postflare loop.

\subsection{The period ratio $P_1/P_2$ of fundamental and first harmonics}

The ratio between the periods of the fundamental and the first
harmonics of sausage waves is proportional to $P_1/P_2 \sim$ 1.68,
which is significantly shifted from 2. Similar phenomena were
observed for fast kink oscillations (Verwichte et al. 2004; De
Moortel \& Brady 2007; Van Doorsselaere et al. 2007). The deviation
of $P_1/P_2$ from 2 in homogeneous loops is very small due to the
wave dispersion (McEwan et al. 2006). But longitudinal density
stratification causes significant shift of $P_1/P_2$ from 2 (Andries
et al. 2005; McEwan et al. 2006; Van Doorsselaere et al. 2007).

Using the results of Andries et al. (2005), we estimate the density
stratification along the loop as $L/\pi H \approx$ 1.8, where $H$ is
the density scale height. The calculation of McEwan et al. (2006)
gives the approximately same value. Thus, both calculation yields
the same density scale height as $\sim$ 17 Mm for the loop length of
97 Mm.

It must be mentioned that seismologically estimated scale height of
17 Mm is larger comparing to the hydrostatical scale height at
H-alpha temperature. This can be explained by two reasons: First, it
is possible that the medium is not in equilibrium inside postflare
loops; Second, some mechanism (e.g., the variation of loop cross
section with height) other than the density stratification causes
the deviation of $P_1/P_2$ from 2. Recently, Verth and
Erd\'elyi (2008) have studied the effect of magnetic stratification
on loop transverse kink oscillations and found that the loop
divergence may have the significant effect on the period ratio
almost similar to the density stratification. It would be
interesting to study the effect of magnetic stratification on fast
sausage oscillations. However, even if the loop divergence effect
reduces the estimated scale height to $\sim$ 7-8 Mm, it still
remains larger than equilibrium scale height. Therefore, we suggest
that the post flare loops are not in equilibrium, which may cause
plasma motions along the loop. In the movie of the
observations, we clearly see the unidirectional mass motion
from western foot point to eastern foot point in the selected loop which 
may indicate its departure from hydrostatic equilibrium. However, it will be
interesting to search further the observational evidence in
high resolution space data. The question is opened for further
discussion.

\subsection{Trapped or leaky mode?}

It is not entirely clear from wavelet analysis that whether the
global oscillation represents trapped or leaky mode. We see in Fig.
2 that the oscillation persists at least for 45 min, i.e., for
$\sim$ 4-5 wave periods. However, the relative intensity
plot in Fig. 2 (upper panel) likely shows that the oscillation
amplitude decreases after 35 min. As intensity oscillations are due
to the variation in plasma density, hence the decrement may reflect
the process of wave damping. Since the oscillations are in sausage
mode, the resonant absorption is unlikely to occur. Therefore, the
wave leakage in the surroundings, is the most probable candidate for
the wave damping. If the oscillation represents the global leaky
mode instead of trapped one, then its phase speed would be close to
the external Alfv\'en speed (Pascoe et al. 2007), which enables us
to estimate its value as $v_{Ae}=2L/P_1 \sim 330$ km/s. However this
estimation is done for the straight magnetic cylinder, while noting
that the curvature probably enhances the wave leakage (Verwichte et
al. 2005; D\'iaz et al. 2006; Selwa et al. 2007).

For comparison, we may estimate the external Alfv\'en speed from the
wave leakage in curved magnetic slab (D\'iaz et al. 2006). Using
equation (25) of D\'iaz et al. (2006), we may write the expression
for the external Alfv\'en speed as
$$
v_{Ae} \sim {{2a}\over {\pi^2 \tau_d}} {n_0\over n_e},
$$
where $\tau_d$ is the damping time. From Fig. 2 , we estimate the
damping time as
$$
\tau_d \sim 20\,\, min,
$$
then using the density ratio of $n_0/n_e \sim 600$ we estimate the
value of the external Alfv\'en speed as $v_{Ae} \sim 300$ km/s
(note, that this estimation is done for a curved slab,
therefore Eq. (3), which is obtained for a straight loop, is not
valid here).

\section{Discussion and Conclusions \label{sec:disc}}

Using high spatial (1.3 arc sec) and temporal (5 and 10 s)
resolution H-alpha observations, we found intensity oscillations with different periodicity
in different parts of cool postflare loop. The intensity shows the
oscillation with the period of $\approx$ 587 s at the loop apex and
with the period of $\approx$ 349 s near loop footpoint. We interpret
the oscillations as signature of the fundamental and first harmonics
of fast sausage mode. It is difficult to say whether the
oscillations are due to the trapped or leaky modes. However,
intensity plot likely shows the decrement of the oscillation
amplitude, therefore we suggest the presence of leaky nature of
these modes. Seismologically estimated Alfv\'en speed outside the
loop is $\sim$ 300-330 km/s. Using the period ratio $P_1/P_2$, we
also estimate the density scale height in the loop as $\sim$ 17 Mm.
This value is much higher than the equilibrium scale height
at low H-alpha temperature, therefore we suggest that cool postflare
loops are not in hydrostatic equilibrium. In the movie of the
observations, we clearly see the unidirectional mass motion
from western foot point to eastern foot point in the selected loop which 
indicates its departure from hydrostatic equilibrium.

In conclusions, we report the first observational evidence of
multiple oscillations of fast sausage modes in cool chromospheric
postflare loop. Future detailed observational search should be
investigated, especially using recent space-based observations.

\section*{Acknowledgments}
AKS wishes to express his gratitude to Prof. Ram Sagar (Director, ARIES) for
valuable suggestions and encouragements,
and Dr. E. O'Shea for providing wavelet tool. TVZ acknowledges the
Georgian National Science Foundation for the
financial support through its grant GNSF/ST06/4-098.We wish to express
our gratitude to the anonymous referee for his valuable comments which considerably
improved our manuscript.

\bsp

\label{lastpage}

\end{document}